\documentstyle[preprint,tighten,floats,prd,aps]{revtex}
\input epsf.tex
\def\DESepsf(#1 width #2){\epsfxsize=#2 \epsfbox{#1}}
\begin{document}
\preprint{\vbox{\hbox{OITS-555}}  }
\draft
\title{Penguin $B$ Decays $b\rightarrow s l^+l^-$ and $b\rightarrow s g$
\footnote{Work supported in part by the Department of Energy Grant No.
DE-FG06-85ER40224.\\
Invited talk presented at the 27th International Conference on High Energy
Physics, Glasgow, Scotland.}}
\author{N.G. Deshpande\footnote{E-mail: desh@oregon.uoregon.edu} and Xiao-Gang
He\footnote{E-mail: he@quark.uoregon.edu}}
\address{Institute of Theoretical Science\\
University of Oregon\\
Eugene, OR 97403-5203, USA}

\date{September 1994}
\maketitle
\begin{abstract}
The penguin mediated processes $b\rightarrow s g$ and $b\rightarrow
s l^+l^-$ are studied. In the Standard Model, for the leptonic modes
improvement in experimental limits will put strigient bounds on the top mass,
where the present limit from $b\rightarrow s \mu^+\mu^-$ is 390 GeV. For
hadronic penguin processes, although the gluonic penguin dominates, we find the
electroweak contribution are around 30\% for the upper range allowed top mass.
The branching ratio for $B\rightarrow X_s \phi$ is predicted to be in the range
$(0.6\sim 2)\times 10^{-4}$. Effects of the charged Higgs in two Higgs doublet
models are discussed.
\end{abstract}
\pacs{}

Rare $B$ decays, particularly pure penguin decays, have been subject of
considerable theoretical and experimental
interest recently\cite{bdecay}. The photonic
penguin induced process $B\rightarrow K^* \gamma$ has been observed
by CLEO collaboration\cite{cleo} and is consistent with the
Standard Model (SM) prediction\cite{desh1}. In this talk we will concentrate
on two
other classes of penguin decays, $b\rightarrow s l^+l^-$ and $b\rightarrow s
g$.
\\
\\
\noindent{\bf Process $b\rightarrow s l^+l^-$}
\\
\\
The process $b\rightarrow s l^+l^-$ is sensitive to top mass unlike
$b\rightarrow s \gamma$,
and improvement in the experimental bound should greatly improve the top
quark mass upper limit which is at present at $\sim$ 390 GeV from $b\rightarrow
s \mu^+\mu^-$\cite{ldesh}. This process for large top mass has dominant
contribution from Z
exchange and the box diagram\cite{soni}.

The effective Hamiltonian density relevant for $b\rightarrow s l^+l^-$ decay
is:
  \begin{eqnarray}
H_{eff} \cong {4 G_F \over \sqrt{2}} \left( V_{cb}V^*_{cs}
\right) \sum \tilde c_j(m) \tilde O_j(m)\;.
  \end{eqnarray}
The important operators for us are:
  \begin{eqnarray}
\tilde O_7 &=& \left(e/16\pi ^2 \right) m_b \left(\overline{s}_L \sigma _{\mu
\nu}  b_R  \right)F^{\mu\nu} \quad ,\nonumber\\
\tilde O_9 &=& \left(e^2/16\pi ^2 \right) \left(\overline{s}_L \gamma _{\mu}
b_L \right) \overline{\ell} \gamma^{\mu}\ell \quad ,\nonumber\\
\tilde O_{10} &=& \left(e^2/16\pi ^2 \right) \left(\overline{s}_L \gamma _{\mu}
b_L \right) \overline{\ell} \gamma^{\mu}\gamma _5\ell\; .
\end{eqnarray}
Here $F_{\mu \nu}$ is the electromagnetic interaction field strength tensor.

The QCD-renomalized coefficients $\tilde c_j(m)$ are calculated in Ref.
\cite{lgsw}, and their implications are discussed in Ref.\cite{ldesh}.
The branching ratio of $b\rightarrow s l^+\l^-$ can be
written after normalizing the rate to $BR(b \rightarrow c e
\overline{\nu}) \approx 0.108$, \cite{lgsw,ldesh1}:
\begin{eqnarray}
&BR( b\rightarrow s l^+\l^-) = \nonumber\\
&K [F_1 (|\tilde c_9|^2
+
|\tilde c_{10}|^2) + F_3 \tilde c_9 \tilde c_7 + F_2 |\tilde c_7|^2 ]\; ,
\end{eqnarray}
where
\begin{eqnarray}
K = \left(\alpha / 4\pi \right)^2 \left(2/\lambda \widetilde{\rho}
\right)BR \left(b \rightarrow ce\overline{\nu} \right) = 1.6 \cdot 10^{-7}
\end{eqnarray}
and $\alpha$ is fine structure constant. The phase space factor
$\widetilde{\rho}$ and the $QCD$ correction factor $\lambda$ for the
semileptonic process are well known \cite{lbarger}.  We have used
$\widetilde\rho = 0.5$ and $\lambda = 0.889$.  The phase space
integration from $min = \left( 2 m_{\ell}/m_b  \right)^2$ to $max =
\left( 1 - m_s/m_b  \right)^2$ give the following values \cite{ldesh2}
for the constants $F_i$:
\begin{eqnarray}
F_1 = 1,\quad F_3 = 8,\quad \mbox{for} \quad min \cong 0,\quad max \cong 1
\quad ,
\end{eqnarray}
\begin{eqnarray}
F_2 = 32 \left[ \ell n \left(m_b/2m_{\ell}\right) \right]; \quad
\mbox{for}\quad max \cong 1; \quad  \ell = e, \mu \,.
\end{eqnarray}

We plot in Fig. 1 the branching ratios for
$b\rightarrow s l^+l^-$ as a function of the top mass for the standard model.
In the SM the process $b\rightarrow s e^+e^-$ is enhanced over $b\rightarrow
s\mu^+\mu^-$ by $\sim 60\%$ for $m_t = 150$ GeV due to the small electron mass
\cite{ldesh2}. As noted, the decay rate is highly dependent on $m_t$ and
improvement on this limit should improve bounds on $m_t$ significantly.

\begin{figure}[htb]
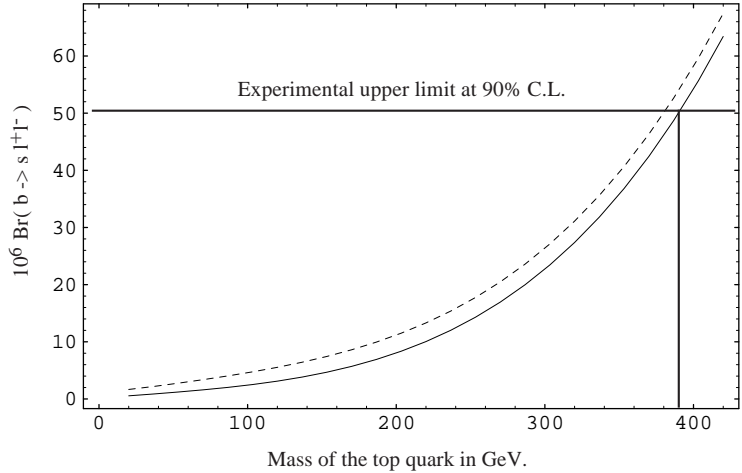

\centerline{ \DESepsf(ichep1.epsf width 10 cm) }
\smallskip
\caption{Branching ratios for $b\rightarrow s e^+e^-$ (dashed line) and
$b\rightarrow s\mu^+\mu^-$ (solid line) as a function of the top mass.}
\label{ichep1}
\end{figure}

In Ref.\cite{ldesh} the implications of additional Higgs doublet on
$b\rightarrow s l^+l^-$
is discussed. The conclusions are:\\
\\
\noindent
1)  In two-Higgs doublet model II, where couplings are like the
minimal supersymmetric model, constraints are possible for all
values of $\tan \beta = v_2/v_1$ for smaller Higgs masses; although the
experimental bounds will have to be improved to draw useful conclusions.\\
\\
\noindent
2)  In the two-Higgs doublet model I, the constraints on Higgs
masses are similar to $b\rightarrow s \gamma$, and tight constraints can be
derived
on Higgs masses only for small values of $\tan \beta = v_2/v_1$.\\
\\
\noindent{\bf Process $b\rightarrow s g$}\\
\\
The gluonic penguin induced $B$ decays are expected
to be observed very soon. A large number of gluonic penguin induced $B$ decay
channels were
studied in Ref.\cite{desh2} using  $\Delta B = 1$ effective Hamiltonian
$H_{\Delta B=1}$ in the lowest nonvanishing order. In Ref.\cite{gatto1} the
next-to-leading order QCD corrected pure gluonic penguin $H_{\Delta B=1}$
was used with top quark mass $m_t$ fixed at
150 GeV. In this talk we present a study of the next-to-leading order QCD
corrected
Hamiltonian $H_{\Delta B = 1}$ in the SM and in two Higgs doublet models,
taking particular care to {\em include the full electroweak contributions} and
find the dependence on $m_t$ and $\alpha_s$. The cleanest signature of hadronic
penguin
processes are: $B\rightarrow X_s \phi$, $B\rightarrow K\phi (K^*\phi)$, and
$B_s\rightarrow \phi\phi$. The process $B\rightarrow X_s\phi$ is particularly
recommended because it is free from form factor uncertainties. We find not only
that the QCD correction in next-to-leading order are large, but also inclusion
of the full electroweak contributions have significant effect on the branching
ratio which could reduce the pure gluonic penguin contribution by 30\% at the
upper range of allowed top quark mass. Our results which have been derived
independently\cite{he}, agree with Fleischer\cite{fleischer}.
The electroweak corrections alter the isospin structure of penguins, and have a
major impact on the analysis of certain $B$ decays. This will be presented in a
forthcoming publication\cite{he1}.

The QCD corrected $H_{\Delta B = 1}$ relevant to us
can be written as follows\cite{laut}:
\begin{eqnarray}
&H_{\Delta B=1} = {G_F\over \sqrt{2}}[V_{ub}V^*_{us}(c_1O^u_1 + c_2
O^u_2)\nonumber\\
&+V_{cb}V^*_{cs}(c_1O^c_1+c_2O^c_2) - V_{tb}V^*_{ts}\sum c_iO_i] +H.C.\;,
\end{eqnarray}
where the Wilson coefficients (WCs) $c_i$ are defined at the scale of $\mu
\approx
m_b$; and $O_i$ are defined as
\begin{eqnarray}
O^q_1 &=& \bar s_\alpha \gamma_\mu(1-\gamma_5)q_\beta\bar
q_\beta\gamma^\mu(1-\gamma_5)b_\alpha\;,\nonumber\\
O^q_2 &=&\bar s \gamma_\mu(1-\gamma_5)q\bar
q\gamma^\mu(1-\gamma_5)b\;,\nonumber\\
O_3 &=&\bar s \gamma_\mu(1-\gamma_5)b
\bar q' \gamma_\mu(1-\gamma_5) q'\;,\nonumber\\
Q_4 &=& \bar s_\alpha \gamma_\mu(1-\gamma_5)b_\beta
\bar q'_\beta \gamma_\mu(1-\gamma_5) q'_\alpha\;,\\
O_5 &=&\bar s \gamma_\mu(1-\gamma_5)b  \bar q'
\gamma^\mu(1+\gamma_5)q'\;,\nonumber\\
Q_6 &=& \bar s_\alpha \gamma_\mu(1-\gamma_5)b_\beta
\bar q'_\beta \gamma_\mu(1+\gamma_5) q'_\alpha\;,\nonumber\\
O_7 &=& {3\over 2}\bar s \gamma_\mu(1-\gamma_5)b  e_{q'}\bar q'
\gamma^\mu(1+\gamma_5)q'\;,\nonumber\\
Q_8 &=& {3\over 2}\bar s_\alpha \gamma_\mu(1-\gamma_5)b_\beta
e_{q'}\bar q'_\beta \gamma_\mu(1+\gamma_5) q'_\alpha\;,\nonumber\\
O_9 &=& {3\over 2}\bar s \gamma_\mu(1-\gamma_5)b  e_{q'}\bar q'
\gamma^\mu(1-\gamma_5)q'\;,\nonumber\\
Q_{10} &=& {3\over 2}\bar s_\alpha \gamma_\mu(1-\gamma_5)b_\beta
e_{q'}\bar q'_\beta \gamma_\mu(1-\gamma_5) q'_\alpha\;.\nonumber
\end{eqnarray}
Here $q'$ is summed over u, d, and s.

We work with renormalization scheme independent WCs $c_i$ as discussed in
Ref.\cite{laut} . In Table 1, we show some sample values of
$c_i$ for some values of $m_t$ with the central value $\alpha_s(m_Z) = 0.118$
and $\mu = m_b$\cite{he}.

\begin{table}

\begin{tabular}{|c|c|c|c|c|c|}
$m_t$(GeV) & $ c_1$ &$ c_2$ &$ c_3$ &$
c_4$&$ c_5$ \\ \hline
130&-0.313&1.150&0.017&-0.037&0.010\\ \hline
174&-0.313&1.150&0.017&-0.037&0.010\\ \hline
210&-0.313&1.150&0.018&-0.038&0.010\\ \hline
$m_t$(GeV)&$ c_6$&$ c_7/\alpha_{em}$&$ c_8/\alpha_{em}$&
$ c_9/\alpha_{em}$&$ c_{10}/\alpha_{em}$\\ \hline
130&-0.045&-0.061&0.029&-0.978&0.191\\ \hline
174&-0.046&-0.001&0.049&-1.321&0.267\\ \hline
210&-0.046&0.060 &0.069&-1.626&0.334
\end{tabular}
\caption{The Wilson coefficients for $\Delta B = 1$ at $m_b = 5\; GeV$ with
$\alpha_s(m_Z) = 0.118$.}
\end{table}

We also need to treat the matrix elements to one-loop level for consistency.
These one-loop matrix elements can be rewritten in terms of the tree-level
matrix elements $<O_j>^{t}$ of the effective operators, and one finds
\cite{fleischer,palmer} $<c_iQ_i>$ to be equal to
\begin{eqnarray} c_i [\delta_{ij}+{\alpha_s\over 4\pi}m^s_{ij}
+{\alpha_{em}\over 4\pi}m^e_{ij}] <O_j>^{t}
\equiv c_i^{eff}<O_i>^{t}.\nonumber\\
\end{eqnarray}
We have worked out the full matrices $m^{s,e}$. For the processes we are
considering only $ c_{3-10}$ contribute.
These are given by,
\begin{eqnarray}
c^{eff}_3 &=& c_3 - P_s/3\;,\;\; c^{eff}_4 = c_4 +P_s\;,\nonumber\\
c^{eff}_5 &=& c_5 - P_s/3\;,\;\;c^{eff}_6 = c_6 + P_s\;,\nonumber\\
c^{eff}_7 &=& c_7 +P_e\;,\;\;\;\;\;\;c_8^{eff} =
c_8\;,\nonumber\\
c_9^{eff} &=&  c_9 +P_e\;,\;\;\;\;\; c_{10}^{eff} =  c_{10}\;.
\end{eqnarray}
The leading contributions to $P_{s,e}$ are given by:
 $P_s = (\alpha_s/8\pi) c_2 (10/9 +G(m_c,\mu,q^2))$ and
$P_e = (\alpha_{em}/9\pi)(3 c_1+ c_2) (10/9 + G(m_c,\mu,q^2))$. Here
$m_c$ is the charm quark mass which we take to be 1.35 GeV. The function
$G(m,\mu,q^2)$ is give by
\begin{eqnarray}
G(m,\mu,q^2) = 4\int^1_0 x(1-x) \mbox{d}x \mbox{ln}{m^2-x(1-x)q^2\over
\mu^2}\;.
\end{eqnarray}
In the numerical calculation, we will use $q^2 = m_b^2/2$ which represents the
average value.

We obtain the
decay amplitude for $B\rightarrow X_s \phi$
\begin{eqnarray}
&A(B \rightarrow X_s \phi) \approx A(b\rightarrow s \phi)=\nonumber\\
&- {g_\phi G_F\over \sqrt{2}}V_{tb}V^*_{ts}\epsilon^\mu
C \bar s \gamma_\mu (1-\gamma_5)b\;,
\end{eqnarray}
where $\epsilon^\mu$ is the polarization
of the $\phi$ particle; $C = c^{eff}_3+c^{eff}_4+c^{eff}_5+\xi
(c^{eff}_3+c^{eff}_4+c^{eff}_6) - (c^{eff}_7+c^{eff}_9+c^{eff}_{10}
+\xi(c^{eff}_8 +c^{eff}_9+c^{eff}_{10}))/2$ with $\xi=1/N_c$,
where $N_c$ is the number of colors.
The coupling constant $g_\phi$ is defined
by $<\phi|\bar s\gamma^\mu s|0> = i g_\phi \epsilon^\mu$. From the experimental
value for $Br(\phi \rightarrow e^+e^-)$, we obtain
$g^2_\phi = 0.0586\; GeV^4$.

The decay rate is, then, given by
\begin{eqnarray}
&\Gamma(B\rightarrow X_s \phi) = {G_F^2g_\phi^2m_b^3 \over 16\pi m_\phi^2}
|V_{tb}V^*_{ts}|^2|C|^2\lambda_{s\phi}^{3/2}\nonumber\\
&\times[1 + {3\over \lambda_{s\phi}}
{m_\phi^2\over m_b^2}(1-{m_\phi^2\over m_b^2} +{m_s^2\over m_b^2})]\;,
\end{eqnarray}
where $\lambda_{ij} = (1-m_j^2/m_b^2 -m_i^2/m_b^2)^2 - 4m_i^2m_j^2/m_b^4$.

We normalize the
branching ratio to the semi-leptonic decay of $B\rightarrow X_c e \bar \nu_e$.
We have
\begin{eqnarray}
&Br(B\rightarrow X_s\phi) = Br(B\rightarrow X_c e \bar \nu_e)
{|V_{tb}V^*_{ts}|^2\over |V_{cb}|^2}\nonumber\\
&\times{12\pi^2g_\phi^2 \lambda_{s\phi}^{3/2}
\over m_\phi^2m_b^2\lambda \tilde \rho} |C|^2[ 1 + {3\over \lambda_{s\phi}}
{m_\phi^2\over m_b^2}(1-{m_\phi^2\over m_b^2} +{m_s^2\over m_b^2})]\;.
\end{eqnarray}

We show in, Fuigure 2 and 3 the predictions for
the branching ratio $Br(B\rightarrow X_s \phi)$ in the SM as a function of top
quark mass $m_t$ and the strong coupling constant $\alpha_s(m_Z)$ with and
without electroweak corrections, and for $N_c = 2$ and 3.

The dominant contribuitons are from the gluonic penguin. There is a very small
$m_t$ dependence for the branching ratio calculated without the inclusion of
the electroweak penguin contributions. The inclusion
of the full electroweak contribuitons have sizeable effects which reduce
the branching ratios by
about 20\% to 30\% for the central value of $\alpha_s$ with $m_t$ varying
from 100 GeV to 200 GeV.
It is clear from Figure 2 and 3 that the full contribution has a large $m_t$
dependence. There may be corrections to the branching ratios predicted
by the factorization method. It is a common practice to parameterize the
possible new contributions by treating  $\xi$ as a free
parameter\cite{buar,gatto2,desh3}. Using experimental values from non-leptonic
$B$ decays, it is found that\cite{gatto2},  $a_1=c_2 +\xi c_1$ and $a_2 =
c_1+\xi c_2$ have the same signs, and $|a_2|\approx 0.27$ and $|a_1| \approx
1.0$. The branching ratios for $N_c = 2$ are
about 2 times those for $N_c = 3$.

\begin{figure}[htb]
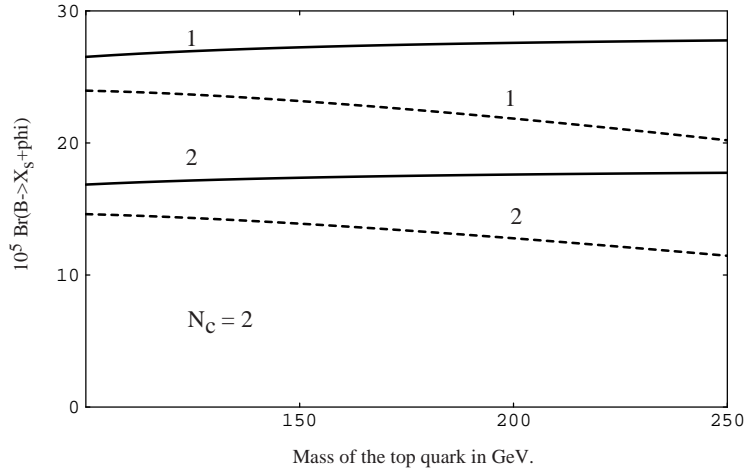

\centerline{ \DESepsf(ichep2.epsf width 10 cm) }
\smallskip
\caption{$Br(B\rightarrow X_s \phi)$ as a function of top mass with
$N_c = 2$, and
$\alpha_s(m_Z) = 0.125$ (curves 1) and $\alpha(m_Z) = 0.111$ (curves 2).
The dashed and solid lines are for the branching ratios with the full strong
and electroweak penguin contributions, and without the electroweak
contributions, respectively.}
\label{ichep2}
\end{figure}

\begin{figure}[htb]
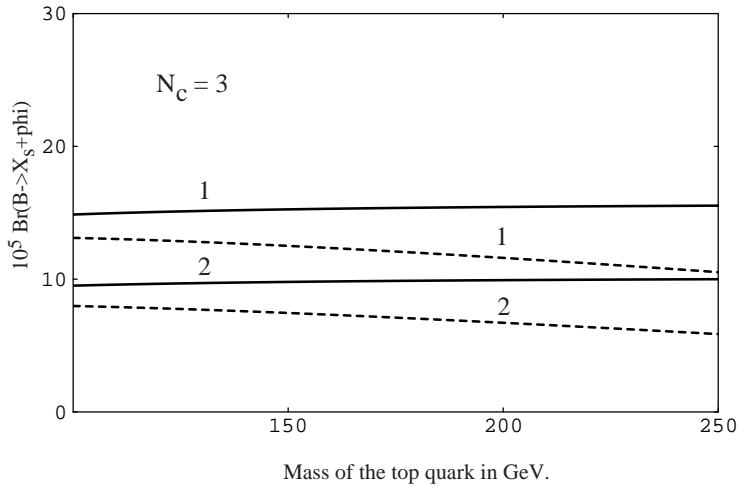

\centerline{ \DESepsf(ichep3.epsf width 10 cm) }
\smallskip
\caption{The same as FIG.2 but with $N_c = 3$.}
\label{ichep3}
\end{figure}

For the central value of $\alpha_s(m_Z)$ and the central value of $m_t = 174$
GeV reported by CDF\cite{cdf}, the value for
$Br(B\rightarrow X_s \phi)$ is about $ 1.7\times 10^{-4}$ for $N_c = 2$.

Using form factors from Refs.\cite{buar,gatto2}, we also calculated the
exclusive decay
rates for $B\rightarrow K\phi$, $B\rightarrow K^*\phi$, and $B_s\rightarrow
\phi\phi$.
The exclusive branching ratios $B\rightarrow
K\phi$ and $B\rightarrow K^*\phi$ are
about the same which are $1\times 10^{-5}$ if the form factors from
Ref.\cite{buar} are used. If the form factors from Ref.\cite{gatto2} are used,
one obtains
 $Br(B\rightarrow K\phi)\approx 1.7\times 10^{-5}$,
 $Br(B\rightarrow K^*\phi)\approx 0.5\times 10^{-5}$, and
$Br(B_s\rightarrow \phi\phi)\approx 0.4 \times 10^{-5}$.

We have also looked at $b\rightarrow s g$ in two-Higgs doublet model\cite{he}.
Both models I and II give the same results.
The ratio of decay rates of the SM predictions to the two-Higgs doublet model
predictions weakly depends on $N_c$. We find that the effects of the charged
Higgs boson contributions are small for
$cot\beta < 1$. When increasing $cot\beta$, the charged
Higgs contributions become important and the effect is to cancel the SM
contributions. When $cot\beta$ becomes very large the charged Higgs boson
contributions become the dominant ones. However, using the information from
$B\rightarrow X_s \gamma$, it is found that for
small $m_H\sim 100$ GeV and $m_t \sim 174$ GeV, $cot\beta$ is constrained to
be less than 1\cite{hh2}. For these values, the charged Higgs boson effects on
the processes discussed in this paper are less than 10\%. For $m_H \sim 500$
GeV, the charged Higgs boson effects can reduce the hadronic penguin $B$
decays by 40\% because the range of $cot\beta$ allowed from $b\rightarrow
s\gamma$ is now larger\cite{hh2}. The effects become smaller for larger
$m_H$.\\
\\
\noindent{\bf $B\rightarrow K\pi/\pi\pi$ modes}\\
\\
We now present a summary of contribution by Hayashi, Joshi, Matsuda and
Tanimoto\cite{joshi}
to this conference. They focus on gluonic effects in the exclusive channels
$B\rightarrow K\pi$ and $B\rightarrow \pi\pi$. These channels have contrbutions
from both the tree operastors $O_{1,2}$ and the penguin operators. their
Hamiltonian includes leading order QCD correction, but not the Z,$\gamma$
penguin and W box contributions to the penguin diagrams. Effect of charm loop
is included in the same manor as discussed by us where $c^{eff}_i$ are
introduced. The value of $q^2$ is taken as $m^2_b/2$ in $P_s$. The
factorization hypothesis is employed, with further assumption that $N_c = 3$,
which might lead to incorrect estimates.

The amplitude for $B^0\rightarrow K^+\pi^-$ and $B^0\rightarrow \pi^+\pi^-$ can
both be expressed in terms of a universal form factor $F^{B\pi}_0(q^2)$ if $B$
annihilation terms are neglected
\begin{eqnarray}
q^\mu < \pi^-|\bar u \gamma_\mu b|B^0_d> = (m_B^2-m_\pi^2) F^{B\pi}_0(q^2)\;.
\end{eqnarray}
This form factor drops out when ratios of $B^0\rightarrow K^+\pi^-$ to
$B^0\rightarrow \pi^+\pi^-$ decay rates are taken. This ratio is, however,
sensitive to $|V_{ub}/V_{cb}|$ and the phase of $V_{ub} = |V_{ub}|e^{-i\phi}$.
The authors find $R_B = \Gamma(B_b^0\rightarrow
K^-\pi^+)/\Gamma(B^0_d\rightarrow \pi^+\pi^-)$ can range from 0.4 to 7.0. They
also calculated the relative contribution of penguin and tree contributions to
$B\rightarrow K\pi$ and $B\rightarrow \pi\pi$ processes. Their results for
ratio of amplitudes for $\phi = 90^0$ are
\begin{eqnarray}
{A(penguin)\over A(tree)} = \left \{
\begin{array}{ll}
4.22{0.08\over |V_{ub}/V_{cb}|}\;, & \;\;\mbox{for} B\rightarrow K\pi\\
0.22{0.08\over |V_{ub}/V_{cb}|}\;, &\;\; \mbox{for} B\rightarrow \pi\pi
\end{array}
\right.
\end{eqnarray}

The present CLEO observation of $BR(B\rightarrow K^+\pi^- + \pi^+\pi^-)
=(2.4^{+0.8}_{-0.7}\pm0.2)\times 10^{-5}$ imposes the limit
$F^{B\pi}_0(0) = 0.26$ to $0.55$ which is consistent with BSW model
$F^{B\pi}_0(0) = 0.33$.

The authors have also considered CP asymmetry in $B\rightarrow \pi^+\pi^-$
decay arising from the phase of $V_{ub}$ as well as the imaginary part
$c^{eff}$.
The asymmetry can be as large as 30\%.

\end{document}